\documentstyle[preprint,tighten,eqsecnum,floats,epsfig,aps]{revtex}

\newcommand{\beq}{\begin{equation}}
\newcommand{\eeq}{\end{equation}}
\newcommand{\mc}{\multicolumn}

\begin{document}
\draft

\title{Weak decays of medium and heavy ${\bf \Lambda}$-hypernuclei}
\author{W. M. Alberico, A. De Pace, G. Garbarino}
\address{Dipartimento di Fisica Teorica, Universit\'a di Torino \\
and INFN, Sezione di Torino, 10125 Torino, ITALY}
\author{A. Ramos}
\address{Departament d'Estructura i Constituents de la Mat\`eria, 
Universitat de Barcelona,\\ 08028 Barcelona, SPAIN}

\date{\today}
\maketitle

\begin{abstract}
We have made a new evaluation of the $\Lambda$ decay width in nuclear
matter
within the Propagator Method. Through the Local Density Approximation it 
is possible to obtain results in finite nuclei. We have also
studied the dependence of the widths on the $NN$ and ${\Lambda}N$ short range
correlations. Using reasonable values for the parameters that control
these correlations, as well as realistic nuclear densities and $\Lambda$ wave
functions,
we reproduce, 
for the first time, the experimental non-mesonic widths in a wide range of mass
numbers (from medium to heavy hypernuclei).
\end{abstract}

\newpage
\pagestyle{plain}
\baselineskip 16pt
\vskip 48pt

\newpage
\section{Introduction}

A hypernucleus is a bound system made of by neutrons, protons and one 
or more hyperons. Among these {\sl strange nuclei}, those which contain 
one ${\Lambda}$ hyperon are the most stable with respect to the 
strong interaction and they are the subject of this paper.
The study of hypernuclear physics may help in understanding some
present
problems related, for instance, to some aspects of weak interactions
in nuclei, or to the origin of the spin-orbit 
interaction in nuclei. Besides, 
it is a good instrument to study the role of quark degrees of freedom in the 
hadron-hadron interactions at short distances and
the renormalization properties of pions in the nuclear medium. 

Nowadays we know some important features of the $YN$ interaction 
\cite{Gi95}. For example,
at intermediate distances the strong ${\Lambda}N$ interaction is weaker than 
the $NN$ one, and its spin-orbit term is very small. 
Moreover, the former has a smaller range than the $NN$ one. 
From the study on mesonic decays of light hypernuclei we have
evidence
for strongly repulsive cores in the ${\Lambda}N$ interaction at short distances
\cite{Ou98},
which automatically appears in quark based models \cite{St90,St93}. 
These characteristics
of the ${\Lambda}N$ interaction are important, as we will see, for the
evaluation of the decay rates of ${\Lambda}$-hypernuclei.

The most interesting hypernuclear decays are those involving weak processes,
which directly concern the hyperon.
The weak decay of hypernuclei occurs via two channels: the so called 
mesonic channel (${\Lambda}\rightarrow \pi N$) and the 
non-mesonic one, in which the pion emitted from the weak hadronic vertex is
absorbed by one or more nucleons in the medium (${\Lambda}N \rightarrow NN$,
${\Lambda}NN \rightarrow NNN$, etc.\ ). Obviously, the non-mesonic
processes can also be mediated by the exchange of more massive mesons than
the pion. The non-mesonic decay is only possible in nuclei and,
nowadays, the study of the
hypernuclear decay is the only practical way to get information on the
weak process
${\Lambda}N \rightarrow NN$, especially on its parity conserving part.
In fact, there are not experimental observations for this interaction using 
lambda beams. It is, however, under study the inverse reaction
$pn\rightarrow {\Lambda}p$ at COSY \cite{Ha95} and RCNP \cite{Ki98}.

The free ${\Lambda}$ decay is
compatible with the ${\Delta}I=1/2$ isospin rule, which is also valid for the
decay of other hyperons and for kaons (namely in non-leptonic
strangeness changing processes). This rule is based on the experimental 
observation 
that the ${\Lambda}\rightarrow \pi^- p$ decay rate is twice the
${\Lambda}\rightarrow \pi^0 n$ one,
but it is not yet understood on theoretical grounds. 
{}From theoretical calculations like the one in ref.~\cite{Os93}
and from experimental measurements \cite{Os98} there is some evidence that the
${\Delta}I=1/2$ rule is broken in nuclear mesonic decay. 
However, this is essentially due to shell
effects and might not be directly related to the weak process. 
A recent estimate of $\Delta I=3/2$ contributions to the
$\Lambda N \to N N$ reaction \cite{Pa98} found  
moderate effects on the hypernuclear decay rates.
In the present calculation of the decay rates in nuclei we will assume this
rule as valid. 
The
momentum of
the final nucleon
in ${\Lambda}\rightarrow \pi N$ is about $100$ MeV for ${\Lambda}$ at rest,
so this process is suppressed by the Pauli principle in nuclei 
(particularly in heavy systems). 
It is strictly forbidden in infinite nuclear matter 
(where $k_F^0\simeq$ 270 MeV), but in finite nuclei it can 
occur because of three important effects: 1) in nuclei the hyperon 
has a momentum distribution that allows larger momenta for the final nucleon,
2) the final pion feels an attraction by the medium such that for fixed
momentum it has a smaller energy than the free one and consequently, due to
energy conservation, the final nucleon again has more chance to come out above 
the Fermi surface, and 3) on the nuclear surface the local Fermi 
momentum is smaller than $k_F^0$ and 
favours the decay. Nevertheless, the mesonic width decreases fastly as 
the mass number $A$ of the hypernucleus increases \cite{Os93}. 
{}From the study of the mesonic channel it could be possible to extract
important information on the pion-nucleus optical potential, which we do not
know today in a complete form. In fact, the mesonic rate is very sensitive to 
the pion self-energy in the medium \cite{Os93}.

The final nucleons in the non-mesonic process
${\Lambda}N\rightarrow NN$ emerge with large
momenta ($\simeq 420$ MeV), so this decay is not 
forbidden by the Pauli principle. On the contrary, apart from very light 
hypernuclei (the $s$-shell ones), 
it dominates over the mesonic decay. The non-mesonic channel is characterized by
large momentum transfers, so that the details of the nuclear structure do not 
have
a substantial influence
while the $NN$ and ${\Lambda}N$ Short Range Correlations 
(SRC) turn out to be very 
important. There is an
anticorrelation between mesonic and non-mesonic decay modes such that the total
lifetime is quite stable from light to heavy hypernuclei \cite{Os98,Co90}: 
${\tau}_{exp}=(0.5\div 1)\, {\tau}_{free}$.

Nowadays, the main problem concerning the weak decay rates 
is to reproduce the experimental values for the ratio ${\Gamma}_n/{\Gamma}_p$
between the neutron and the proton induced widths 
${\Lambda}n\rightarrow nn$ and ${\Lambda}p\rightarrow np$. The theoretical
calculations underestimate the experimental data for all the considered 
hypernuclei \cite{Os98,Pa98,Pa97,Ok98,Du96,It95}:
\beq
\left\{\frac{{\Gamma}_n}{{\Gamma}_p}\right\}^{Th}\ll
\left\{\frac{{\Gamma}_n}{{\Gamma}_p}\right\}^{Exp}
\hspace{0.8in}
0.5\lesssim \left\{\frac{{\Gamma}_n}{{\Gamma}_p}\right\}^{Exp}\lesssim 2 .
\eeq
\noindent In the One Pion Exchange (OPE) approximation the values for this 
ratio are $0.1\div 0.2$. On the other hand the
OPE model has been able to reproduce the 1-body stimulated non-mesonic rates 
${\Gamma}_{1}={\Gamma}_n+{\Gamma}_p$ for light and medium hypernuclei
\cite{Pa97,Ok98,It95}. In order to solve this problem many attempts have
been made up to now, but without success. Among these we recall the 
inclusion in the ${\Lambda}N\rightarrow NN$
transition potential of mesons heavier than the pion 
\cite{Pa97,Du96,It95}, the inclusion of interaction terms that
violate the ${\Delta}I=1/2$ rule \cite{Pa98} and the description of the 
short range baryon-baryon interaction in terms of quark degrees of freedom
\cite{Ok98}. This last calculation 
is the only 
one that has found a consistent (but not sufficient) increase of the neutron 
to proton ratio with respect to the OPE one.
However, this calculation is
only made for $s$-shell hypernuclei and their effective quark-lagrangian 
does not reproduce
the experimental ratio between the ${\Delta}I=1/2$ and ${\Delta}I=3/2$ 
transition amplitudes for the ${\Lambda}$ free decay.

The analysis of the ratio ${\Gamma}_n/{\Gamma}_p$ is influenced by the
2-nucleon induced process ${\Lambda}NN\rightarrow NNN$.
By assuming that the meson produced in the weak
vertex is mainly absorbed by a neutron-proton strongly correlated pair, 
the 3-body process turns out to be
${\Lambda}np\rightarrow nnp$, so that 
a considerable fraction of the measured neutrons 
could come from this channel and not only from the
${\Lambda}n\rightarrow nn$ and ${\Lambda}p\rightarrow np$ ones. 
In this way it might be
possible to explain the large experimental ${\Gamma}_n/{\Gamma}_p$ ratios,
which originally have been analyzed without taking into account 
the 2-body stimulated process. 
Nevertheless, the situation is far from being clear and simple. The new
non-mesonic mode was introduced in ref.~\cite{Al91} and its calculation 
was improved in ref.~\cite{Ra95}, where 
the authors found that the inclusion of the
new channel would lead to larger values of the ${\Gamma}_n/{\Gamma}_p$
ratios extracted from the experiment, 
somehow more in disagreement with theoretical estimates. However, in the
hypothesis that only two nucleons from the 3-body decay are detected, the
reanalysis of the experimental data would lead to smaller ratios \cite{Os98}. 
These observations show that ${\Gamma}_n/{\Gamma}_p$ is sensitive to the 
energy spectra of the emitted nucleons, whose calculation also requires a careful
treatment of the nucleon Final State Interaction.
In ref.~\cite{Ra97} the energy distributions were
calculated using a Monte Carlo simulation to describe the final state interactions. 
A direct comparison of those spectra with 
the experimental ones
favours ${\Gamma}_n/{\Gamma}_p$ values around $2\div 3$ (or higher), in
disagreement with the OPE predictions. However, it was also pointed out the 
convenience of measuring the number of protons per decay event. This observable,
which can be measured from delayed fission events in the decay of heavy
hypernuclei, gives a more reliable ratio
$\Gamma_n/\Gamma_p$ and is less sensitive to details of the Monte Carlo
simulation determining the final shape of the spectra.

In this paper we present a new evaluation of the decay rates for medium to heavy
hypernuclei based on the Propagator Method of ref.~\cite{Os85}, which allows a
unified treatment of all the decay channels. The parameters of
the model are adjusted to reproduce the non-mesonic width of $^{12}_{\Lambda}$C
and the decay
rates of heavier hypernuclei are predicted. We also discuss how the new model
affects the energy spectrum of the emitted nucleons, 
in the hope of obtaining
a ratio $\Gamma_n/\Gamma_p$ more in agreement with the experimental observation. 
 
The paper is organized as follows. In Sec.~\ref{pm} we present 
the model used for
the calculation of the decay rates. 
Our results are presented and discussed in Sec.~\ref{res}. We first study
the sensitivity of the decay rates to the parameters defining the $NN$
and $\Lambda N$ short range
correlations as well as to the nuclear density and $\Lambda$ wave functions.
We then obtain the widths for various hypernuclei and discuss
the energy distribution of the nucleons from the weak decays.
Our conclusions are given in Sec.~\ref{concl}. 
\section{Propagator Method}
\label{pm}
The ${\Lambda}$ decay in nuclear systems can be studied in Random Phase
Approximation (RPA) using the Propagator Method \cite{Ra95,Os85}.
This technique provides a unified picture
of the different decay channels and it is equivalent to the standard
Wave Function
Method (WFM) \cite{Os94}, used by other authors in 
refs.~\cite{Os93,Pa98,Pa97,Du96,It95}. 
The calculation of the widths is usually performed in nuclear matter, and
then extended to finite nuclei via the Local Density Approximation (LDA).
For the calculation of the mesonic rates the WFM is more reliable than the
propagator method in LDA since this channel is rather sensitive to the 
shell structure of the hypernucleus, given the small energies involved. 
Moreover, it is advisable to avoid the use of the LDA for very light 
systems and 
we will make the calculation starting from $^{12}_{\Lambda}$C. 
On the other hand,
the propagator method in LDA offers the possibility of calculations 
over a broad range of mass numbers.

\begin{figure}
\begin{center}
\mbox{\epsfig{file=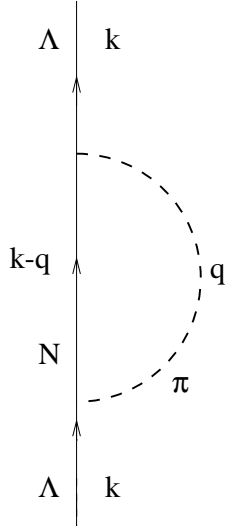}}
\vskip 2mm
\caption{${\Lambda}$ self energy in nuclear matter}
\label{self}
\end{center}
\end{figure}

\begin{figure}
\begin{center}
\mbox{\epsfig{file=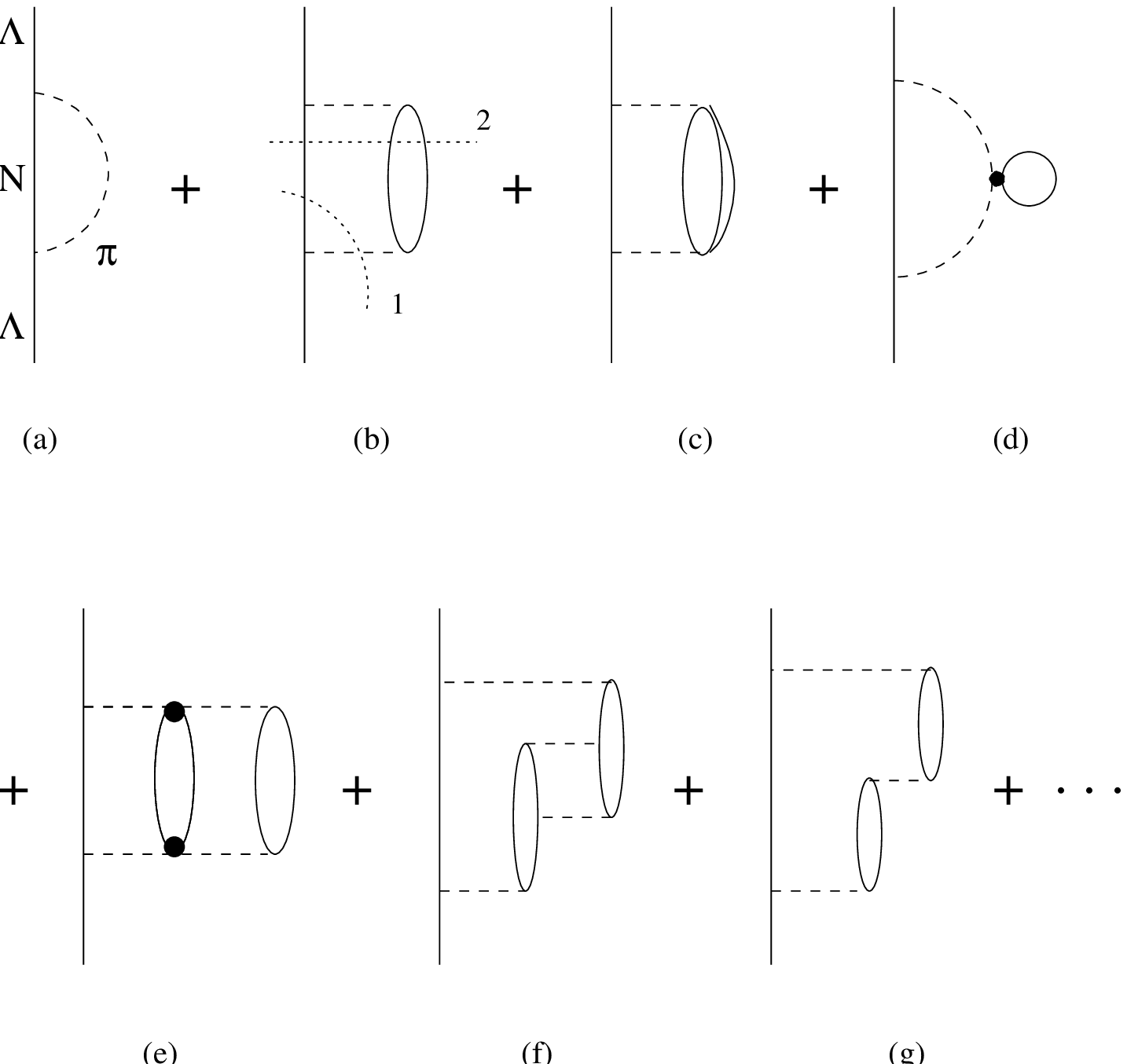,width=.8\textwidth}}
\vskip 2mm
\caption{Lowest order terms for the ${\Lambda}$ self-energy in 
nuclear matter. The meaning of the various diagramms is explained in the text.}
\label{self1}
\end{center}
\end{figure}

The method was introduced in
ref.~\cite{Os85} and we  briefly summarize it here for clarity.
The ${\Lambda}\rightarrow \pi N$ effective lagrangian is:
\beq
\label{lagran}
{\cal L}_{{\Lambda}\pi N}=G m_{\pi}^2\overline{\psi}_N(A+B\gamma_5)
{\bbox \tau} \cdot {\bbox \phi}_{\pi}{\psi}_{\Lambda}+h.c. ,
\eeq
where the values of the weak coupling constants 
$G\simeq 2.211\cdot 10^{-7}/m_{\pi}^2$, $A=1.06$, 
$B=-7.10$ are fixed on the free ${\Lambda}$ decay. The constants $A$ and $B$ 
determine the strengths of the parity
violating and parity conserving ${\Lambda}\rightarrow \pi N$ amplitudes
respectively.
In order to enforce the ${\Delta}I=1/2$ rule, in eq.~(\ref{lagran}) 
the hyperon is assumed to be an isospin spurion with $I_z=-1/2$.
To calculate the ${\Lambda}$ width in nuclear matter we start with the 
imaginary part of the ${\Lambda}$ self-energy:
\beq
\label{Gamma}
{\Gamma}_{\Lambda}=-2{\rm Im}{\Sigma}_{\Lambda} .
\eeq
By the use of Feynman rules, from fig.~\ref{self} it is easy to obtain
the ${\Lambda}$ self-energy in the following form:
\beq
\label{Sigma1}
{\Sigma}_{\Lambda}(k)=3i(G m_{\pi}^2)^2\int \frac{d^4q}{(2\pi)^4}
\left\{S^2+\frac{P^2}{m_{\pi}^2}\bbox q^2\right\}F_{\pi}^2(q)
G_N(k-q)G_{\pi}(q) .
\eeq
Here, $S=A$, $P=m_{\pi}B/2m_N$, while the nucleon and pion 
propagators in nuclear matter are respectively:
\beq
\label{propnucl}
G_N(p)=\frac{{\theta}(\mid \bbox p \mid-k_F)}{p_0-E_N(\bbox p)-V_N+i{\epsilon}}+
\frac{{\theta}(k_F-\mid \bbox p \mid)}{p_0-E_N(\bbox p)-V_N-i{\epsilon}} ,
\eeq
and:
\beq
\label{proppion}
G_{\pi}(q)=\frac{1}{q_0^2-\bbox q^2-m_{\pi}^2-{\Sigma}_{\pi}^*(q)} .
\eeq
In the above, $p=(p_0,\bbox p)$ and $q=(q_0,\bbox q)$ denote 
four-vectors, $k_F$ is the Fermi momentum, $E_N$ is the nucleon total free
energy, $V_N$ is the nucleon binding energy, and ${\Sigma}_{\pi}^*$ 
is the pion proper self-energy in nuclear matter. 
Moreover, in eq.~(\ref{Sigma1}) we have included a monopole form factor for 
the $\pi\Lambda N $ vertex:
\beq
\label{ff}
F_{\pi}(q)=\frac {{\Lambda}_{\pi}^2-m_{\pi}^2}{{\Lambda}_{\pi}^2-q_0^2+\bbox q^2}
\eeq
(the same is used for the $\pi NN$ strong vertex),
with cut-off ${\Lambda}_{\pi}=1.2$ GeV. In fig.~\ref{self1} we show
the lowest order Feynman graphs for the ${\Lambda}$ 
self-energy in nuclear matter. Diagram (a) represents the bare
self-energy term, including the effects of Pauli principle and of
binding on the intermediate nucleon.
In (b) and (c) the pion couples to a particle-hole ({\sl p-h}) and a 
{\sl ${\Delta}$-h} pairs, respectively. Diagram (d) is an insertion of $s$-wave
pion self-energy
at lowest order. In diagram (e) we show a {\sl 2p-2h} excitation coupled to the
pion through $s$-wave $\pi N$ interactions. 
Other {\sl 2p-2h} excitations, coupled in $p$-wave, are 
shown in (f), while (g) is a RPA iteration of diagram (b). 
It is possible to
evaluate the integral over $q_0$ in (\ref{Sigma1}), and the 
${\Lambda}$ self-energy (eq.~(\ref{Gamma})) in nuclear matter becomes 
\cite{Os85}:
\begin{eqnarray}
\label{Sigma2}
{\Gamma}_{\Lambda}(\bbox k,\rho)&=&-6(G m_{\pi}^2)^2\int \frac{d\bbox q}
{(2\pi)^3}{\theta}(\mid \bbox k- \bbox q \mid -k_F)
{\theta}(k_0-E_N(\bbox k-\bbox q)-V_N) \nonumber
\\
& & \times {\rm Im}{\alpha}(q)\mid _{q_0=k_0-E_N(\bbox k-\bbox q)-V_N} ,
\end{eqnarray}
where
\begin{eqnarray}
\label{Alpha}
{\alpha}(q)&=&\left\{S^2+\frac{P^2}{m_{\pi}^2}\bbox q^2\right\}F_{\pi}^2(q)
G_{\pi}^0(q)+\frac{\tilde{S}^2(q)U(q)}{1-V_L(q)U(q)} \nonumber \\
& & +\frac{\tilde{P}_L^2(q)U(q)}{1-V_L(q)U(q)}+
\frac{\tilde{P}_T^2(q)U(q)}{1-V_T(q)U(q)} .
\end{eqnarray}
In eq.~(\ref{Sigma2}) the first ${\theta}$ function forbids
intermediate nucleon momenta (see fig.~\ref{self}) smaller than the 
Fermi momentum and the second one requires the pion energy $q_0$ to be 
positive. Moreover, the ${\Lambda}$ energy, 
$k_0=E_{\Lambda}(\bbox k)+V_{\Lambda}$,
contains a binding term. The pion lines of fig.~\ref{self1} 
have been replaced in eq.~(\ref{Alpha}) by the interactions 
$\tilde{S}$, $\tilde{P}_L$, $\tilde{P}_T$,$V_L$, $V_T$,
which include $\pi$ and $\rho$ exchange modulated by the effect of short range
correlations and 
whose espressions are given in the Appendix.
The functions $V_L$ and $V_T$ represent the (strong) {\sl p-h} interaction,
including a short range
Landau parameter $g^{\prime}$, while $\tilde{S}$, $\tilde{P}_L$ and
$\tilde{P}_T$ correspond to the lines connecting weak and strong 
hadronic vertices and contain another short range Landau parameter 
$g^{\prime}_{\Lambda}$. Furthermore, in eq.~(\ref{Alpha}):
\beq
G_{\pi}^0(q)=\frac{1}{q_0^2-\bbox q^2-m_{\pi}^2} ,
\eeq
is the free pion propagator,
while $U(q)$ contains the Lindhard functions for {\sl p-h} and 
{\sl ${\Delta}$-h} 
excitations \cite{Wa71} and also accounts for {\sl 2p-2h} excitations:
\beq
U(q)=U_{ph}(q)+U_{\Delta h}(q)+U_{2p2h}(q) .
\eeq
It appears in eq.~(\ref{Alpha}) within the standard RPA expression.
Eq.~(\ref{Sigma2}) depends explicitly and through $U(q)$
on the nuclear matter density $\rho=2k_F^3/3{\pi}^2$.
The Lindhard functions $U_{ph}$, $U_{\Delta h}$ are normalized as
in ref.~\cite{Os90} and $U_{2p2h}$ is evaluated as in \cite{Ra95}, that is, by
calculating the available phase space for {\sl 2p-2h} excitations and by taking into
account the experimental data on pionic atoms.
$U(q)$ is related to the pion proper self-energy through:
\beq
{\Sigma}_{\pi}^*(q)={\Sigma}_{\pi}^{(p)\,*}(q)+{\Sigma}_{\pi}^{(s)\,*}(q), 
\hspace{0.5in}
{\Sigma}_{\pi}^{(p)\,*}(q)=\frac{\displaystyle
\frac{f_{\pi}^2}{m_{\pi}^2} \bbox q^2 F_{\pi}^2(q)U(q)}
{1-\displaystyle \frac{f_{\pi}^2}{m_{\pi}^2}g_L(q)U(q)} ,
\eeq
where the Landau function $g_L(q)$ is given in the Appendix [see
eq.~(\ref{gl})],
and
${\Sigma}_{\pi}^{(s)\,*}$ is the $s$-wave part of the self-energy. We will
use the parametrization of ref. \cite{Se83}: 
${\Sigma}_{\pi}^{(s)\,*}(q)=-4\pi (1+m_{\pi}/m_N)b_0\rho$, 
with $b_0=-0.0285/m_{\pi}$. The function ${\Sigma}_{\pi}^{(s)\,*}$ is real
(constant and 
positive), 
therefore it contributes only to the mesonic decay 
(diagram (d) in fig.~\ref{self1} is the relative lowest order). 
On the contrary, the $p$-wave self-energy 
${\Sigma}_{\pi}^{(p)\,*}$ is complex and attractive 
(that is, ${\rm Re}{\Sigma}_{\pi}^{(p)\,*}(q)<0$).

The decay widths in finite nuclei are obtained in LDA. In this
approximation, the Fermi momentum becomes $r$-dependent (that is,
a local Fermi sea of nucleons is introduced) and related again to the
nuclear density by:
\beq
\label{local}
k_F(\bbox r)=\left\{\frac{3}{2}{\pi}^2\rho 
(\bbox r)\right\}^{1/3} .
\eeq
Besides, the nucleon binding potential $V_N$ also becomes $r$-dependent in 
LDA. In Thomas-Fermi approximation we assume:
\beq
\epsilon_F(\bbox r)+V_N(\bbox r)\equiv \frac{k_F^2(\bbox r)}{2m_N}+V_N(\bbox r)=0 .
\eeq
For the ${\Lambda}$ binding energy we use instead the experimental value
\cite{Pi91,Ha96}.
With these prescriptions we can then evaluate the decay width in finite nuclei
through the relation:
\beq
\label{local1}
{\Gamma}_{\Lambda}(\bbox k)=\int d\bbox r \mid {\psi}_{\Lambda}(\bbox r)\mid ^2
{\Gamma} _{\Lambda}[\bbox k,\rho (\bbox r)] ,
\eeq
where ${\psi}_{\Lambda}$ is the ${\Lambda}$ wave function and 
${\Gamma} _{\Lambda}[\bbox k,\rho (\bbox r)]$ is given by 
eqs.~(\ref{Sigma2}), (\ref{Alpha}). This decay rate is valid for fixed 
${\Lambda}$ momentum $\bbox k$. A further average over the ${\Lambda}$ momentum 
distribution gives the total width:
\beq
\label{local2}
{\Gamma}_{\Lambda}=\int d\bbox k \mid \tilde{\psi}_{\Lambda}(\bbox
k)\mid^2{\Gamma}_{\Lambda}
(\bbox k) ,
\eeq
which can be compared with experimental results.

The propagator method provides a unified picture of the decay widths.
The imaginary part of a self-energy diagram requires placing simultaneously
on-shell the particles of the considered intermediate state. For instance,
diagram (b) in fig.~\ref{self1} has two sources of imaginary part. 
One comes from cut 1, where the nucleon and the pion are placed on-shell.
This term contributes to the mesonic channel: the final pion interacts with the
medium through a {\sl p-h} excitation and then escapes from the nucleus.
Diagram (b) and further iterations lead to a
renormalization of the pion in the medium which increases the mesonic rate by
about two orders of magnitude in heavy nuclei \cite{Os85}.
The cut 2 in fig.~\ref{self1}(b) place a nucleon and a {\sl p-h} pair on shell,
so it is the lowest order contribution to the physical process 
${\Lambda}N\rightarrow NN$. 

The mesonic width ${\Gamma}_M$ is calculated from
\beq
\label{alphaprop}
{\alpha}(q)=\left\{S^2+\frac{P^2}{m_{\pi}^2}\bbox q^2\right\}G_{\pi}(q) ,
\eeq
by omitting ${\rm Im}{\Sigma}_{\pi}^*$ in $G_{\pi}$, namely by replacing:
\beq
\label{mesonic}
{\rm Im}G_{\pi}(q)\rightarrow -\pi \delta (q_0^2-\bbox q^2-m_{\pi}^2-
Re{\Sigma}_{\pi}^*(q)) .
\eeq
The one-body induced non-mesonic decay rate ${\Gamma}_1$ 
is obtained by substituting in eqs.~(\ref{Sigma2}), (\ref{Alpha}):
\beq
\label{pres}
{\rm Im} \frac{U(q)}{1-V_{L,T}(q)U(q)}\rightarrow \frac{{\rm Im}U_{ph}(q)}
{\mid 1-V_{L,T}(q)U(q)\mid ^2} ,
\eeq
that is by omitting the imaginary parts of $U_{\Delta h}$ and $U_{2p2h}$ in the
numerator. Indeed
${\rm Im}U_{\Delta h}$ accounts for the ${\Delta}\rightarrow \pi N$ 
decay width, thus representing a contribution to the mesonic decay. There is no
overlap between ${\rm Im}U_{ph}(q)$ and the pole $q_0={\omega}(\bbox q)$ in
eq.~(\ref{mesonic}), so the separation of the mesonic and 2-body 
non-mesonic channels is unambiguous. The renormalized pion pole in 
eq.~(\ref{alphaprop}) is given by the dispersion relation:
\beq
{\omega}^2(\bbox q)-\bbox q^2-m_{\pi}^2-Re{\Sigma}_{\pi}^*
[{\omega}(\bbox q),\bbox q]=0 ,
\eeq
with the constraint:
\beq
q_0=k_0-E_N(\bbox k -\bbox q)-V_N .
\eeq
At the pion pole ${\rm Im}U_{2p2h}\neq 0$, thus the 2-body induced non-mesonic
width ${\Gamma}_2$ 
cannot be calculated using the prescription (\ref{pres}) with $U_{2p2h}$
instead of $U_{ph}$ in the numerator of the r.\,h.\,s. Part of the decay 
rate
calculated in this way it is due to excitations of the renormalized pion and
contributes to ${\Gamma}_M$. The 3-body non-mesonic rate is then calculated by
subtracting ${\Gamma}_M$ and ${\Gamma}_1$ from the total rate ${\Gamma}_{TOT}$, 
which we get
via the full espression for ${\alpha}$ [Eq.~(\ref{Alpha})]. 
\section{Results and discussion}
\label{res}
Let us now discuss the numerical results one can obtain from the above
illustrated formalism. We shall first study the influence of short range
correlations and the $\Lambda$ wave function on the decay width of
$^{12}_\Lambda$C, which will
be used as a testing ground for the theoretical framework in order to fix the
parameters of our model. We will then obtain the decay widths of heavier
hypernuclei and we will explore whether the refined model 
influences the energy distribution of the emitted particles, following the 
Monte Carlo procedure of ref. \cite{Ra97}.

In order to evaluate the width (\ref{local1}) in LDA one needs to specify 
the nuclear density and the wave
function for the ${\Lambda}$. The former is assumed to be a Fermi distribution:
\beq
{\rho}_A(r)=\displaystyle \frac {{\rho}_0(A)}{1+e^{[r-R(A)]/a}} \hspace {0.4 in}
\left[{\rho}_0(A)=\displaystyle \frac {A}{\frac{4}{3} \pi R^3(A)
\{1+[\frac{\pi a}{R(A)}]^2\}}\right] ,
\vspace{0.3 in}
\eeq
\noindent
with radius $R(A)=1.12A^{1/3}-0.86A^{-1/3}$ fm and thickness $a=0.52$ fm.
The ${\Lambda}$ wave function is obtained from a Wood-Saxon (W-S) well which
exactly reproduces the first two single particle eigenvalues ($s$ and $p$ levels)
measured in ${\Lambda}$-hypernuclei. 
\subsection{Short range correlations and 
$\Lambda$ wave function}
\label{resgamma}

A crucial ingredient in the calculation of the decay
widths is the short range part of the $NN$ and ${\Lambda}N$ interactions. They
are expressed by the functions $g_{L,T}(q)$ and $g_{L,T}^{\Lambda}(q)$,
which are reported in the Appendix and contain the Landau parameters $g^{\prime}$ and
$g^{\prime}_{\Lambda}$. No experimental information is available on 
$g^{\prime}_{\Lambda}$, while many constraints have been set on $g^{\prime}$, for
example by the well known quenching of the Gamow-Teller resonance. Realistic
values of $g^{\prime}$, within the framework of the ring approximation, are in the
range $0.6\div 0.7$ \cite{Os82}. However, in the present context $g^{\prime}$
correlates not only {\sl p-h} pairs but also {\sl p-h} with {\sl 2p-2h} states. 
In order to
fix these parameters we shall compare our calculations with the experimental
non-mesonic width of $^{12}_{\Lambda}$C. 

\begin{figure}
\begin{center}
\mbox{\epsfig{file=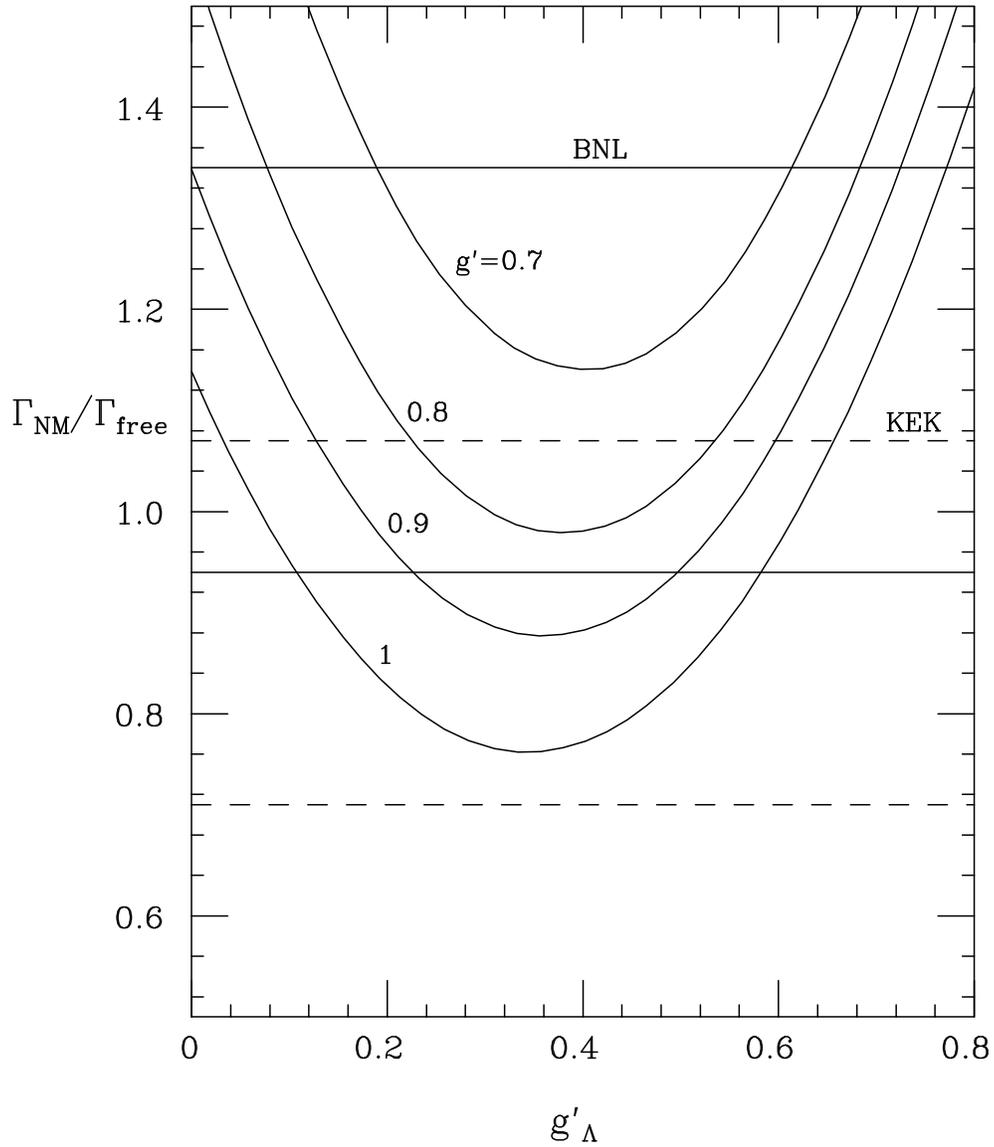,width=.8\textwidth}}
\vskip 2mm
\caption{Dependence of the non-mesonic width on the Landau 
parameters $g^{\prime}$ and $g^{\prime}_{\Lambda}$ for $^{12}_{\Lambda}$C. The experimental
value from BNL~\protect\cite{Sz91} (KEK~\protect\cite{No95}) lies in between the horizontal 
solid (dashed) lines.}
\label{carb}
\end{center}
\end{figure}

In fig.~\ref{carb} we see how the total non-mesonic width for carbon 
depends on
the Landau parameters. 
The rate decreases as $g^{\prime}$ increases. This characteristic
is well established in RPA. Moreover, fixing $g^{\prime}$, 
there is a minimum for $g^{\prime}_{\Lambda}\simeq 0.4$ (almost independent of 
the value of $g^{\prime}$). This is due to the fact that for $g^{\prime}_{\Lambda}\ll 0.4$ the
longitudinal $p$-wave contribution in eq.~(\ref{Alpha})
dominates over the transverse one and 
the opposite occurs for $g^{\prime}_{\Lambda}\gg 0.4$. We also
remind that the $s$-wave interactions are independent 
of $g^{\prime}_{\Lambda}$ [eq.~(\ref{s})]. Moreover, the longitudinal
$p$-wave ${\Lambda}N \rightarrow NN$ interaction [eq.~(\ref{pl})] contains
the pion exchange plus SRC, while the 
transverse $p$-wave ${\Lambda}N \rightarrow NN$ interaction [eq.~(\ref{pt})]
only contains repulsive correlations,
so with increasing $g^{\prime}_{\Lambda}$
the $p$-wave longitudinal contribution to the width decreases, 
while the $p$-wave transverse part increases. From fig.~\ref{carb} 
we see that there is
a broad range of choices of $g^{\prime}$ and $g^{\prime}_{\Lambda}$ values which fit the
experimental band. The latter represents the non-mesonic decay width which is
compatible with both the BNL \cite{Sz91} and KEK \cite{No95} experiments. 
One should notice that the theoretical curves reported in fig.~\ref{carb}
contain the contribution of the 3-body process; should the latter be neglected
(ring approximation) then one could get equivalent results with $g^{\prime}$
values smaller than the ones reported in the figure 
(tipically $\Delta g^{\prime}\simeq -0.1$). The phenomenology of the $(e,e')$ 
quasi-elastic scattering suggests, in ring approximation, $g^{\prime}$ values of
the order of 0.7. Here, by taking into account also $2p-2h$ contributions, we
shall use the ``equivalent" value $g^{\prime}=0.8$, together with $g^{\prime}_{\Lambda}=0.4$.

We note that the values used in ref.~\cite{Ra95}, namely $g^{\prime}=0.615$
and $g^{\prime}_{\Lambda}=0.62$, would yield $\Gamma_1=1.26$ and $\Gamma_2=0.25$, adding
to a non-mesonic width $\Gamma_{NM}=1.51$, which is 50\% larger than the
experimental one. Thus the analysis performed here shows that the present
data for $^{12}_\Lambda$C favour a somewhat different but still 
reasonable $g^{\prime}$ value. 

We shall now illustrate the sensitivity of our calculation to the 
${\Lambda}$ wave
function in $^{12}_{\Lambda}$C. In addition
to the W-S that reproduces the $s$ and $p$ levels, we also use
a harmonic oscillator wave
function with an "empirical" frequency $\omega$ \cite{Pi91,Ha96}, 
again obtained from the $s-p$ energy shift, the 
W-S wave function of Dover {\sl et al.}\ \cite{Do88} and
the microscopic wave function calculated from a non-local self-energy using
a realistic $YN$ interaction in
ref.~\cite{Po98}.
The results are shown in table~\ref{wf sens}, where 
they are compared with the
experimental data from BNL \cite{Sz91} and KEK \cite{No95,Bh98,Sa98}.
\begin{table}[t]
\begin{center}
\caption{Wave function sensitivity} 
\label{wf sens}
\begin{tabular}{c|c c c c c c c}
\mc {1}{c|}{$ ^{12}_{\Lambda}$C} &
\mc {1}{c}{H.O.} &
\mc {1}{c}{Dover} &
\mc {1}{c}{New W-S} &
\mc {1}{c}{Microscopic} &
\mc {1}{c}{BNL \cite{Sz91}} &
\mc {1}{c}{KEK \cite{No95}} &
\mc {1}{c}{KEK New \cite{Bh98,Sa98}} \\ \hline\hline
${\Gamma}_M$     & 0.26 & 0.25 & 0.25 & 0.25 & $0.11\pm 0.27$ & $0.36\pm 0.15$ 
& $>0.11$\\
${\Gamma}_1$     & 0.78 & 0.77 & 0.82 & 0.69 &              &          &    \\
${\Gamma}_2$     & 0.15 & 0.15 & 0.16 & 0.13 &         &          &    \\
${\Gamma}_{NM}$  & 0.93 & 0.92 & 0.98 & 0.81 & $1.14\pm 0.20$ & $0.89\pm 0.18$
& $<1.03$ \\
${\Gamma}_{TOT}$ & 1.19 & 1.17 & 1.23 & 1.06 & $1.25\pm 0.18$ & $1.25\pm 0.18$
& $1.14\pm 0.08$ \\ 
\end{tabular}
\end{center}
\end{table}

By construction, the chosen $g^{\prime}$ and $g^{\prime}_{\Lambda}$ reproduce the 
experimental decay widths using the
W-S wave function which gives the right $s$ and $p$ levels in
$^{12}_\Lambda$C. 
We note that it is possible to generate the microscopic wave function of
ref.~\cite{Po98} 
for carbon via a local hyperon-nucleus W-S potential with 
radius $2.92$ fm and depth $-23$ MeV. 
Although this potential reproduces fairly well the experimental $s$-level for the 
${\Lambda}$ in $^{12}_{\Lambda}$C,  it
does not reproduce the $p$-level.
In this work we prefer to use a completely phenomenological
$\Lambda$-nucleus
potential that can easily
be extended to heavier nuclei and reproduces the experimental
$\Lambda$ single particle levels as well as possible.
Except for $s$-shell hypernuclei, where experimental data require
$\Lambda$-nucleus potentials with a repulsive core
at short distances \cite{Ou98}, the $\Lambda$ binding energies have
been well reproduced by W-S potentials.
We thus use a W-S potential with fixed diffuseness
($a=0.6$ fm) and adjust 
the radius and depth 
to reproduce the $s$ and $p$ $\Lambda$-levels. The paramaters 
of the potential for carbon are $R=2.27$ fm and $V_0=-32$ MeV.

To analyse the results of table \ref{wf sens}, we note that the
microscopic wave function 
is substantially more extended than all the other wave functions used in
the present study.
The Dover parameters \cite{Do88}, namely $R=2.71$ fm, $V_0=-28$ MeV, 
give rise to
a $\Lambda$ wave function that is somewhat more extended than the new W-S one but 
is very similar to that
obtained from a harmonic oscillator with a frequency of $10.9$ MeV, 
adjusted to the $s-p$ energy shift in carbon. 
Consequently, the non-mesonic width from the Dover's
wave function is very similar to the one obtained from the harmonic oscillator
and slightly smaller than the new W-S one. 
The microscopic wave-function predicts the smallest non-mesonic widths due to
the more extended $\Lambda$ wave-function, which explores regions of lower 
density and thus has a smaller probability of interacting 
with one or more
nucleons. 
From table~\ref{wf sens} we also see
that, against intuition, the mesonic width is quite insensitive to 
the ${\Lambda}$ wave function.
On this point we remind that for fixed ${\Lambda}$-momentum, the more
extended is the wave function in $r$-space, the larger is the mesonic
width, since the Pauli blocking effects on the emitted nucleon are reduced.
But, when we make the integral over the ${\Lambda}$-momenta in LDA 
(eq.~(\ref{local2})), to 
more extended wave functions in $r$-space correspond less extended
momentum distributions which tend to decrease the mesonic 
width. The two effects tend to cancel each other and 
${\Gamma}_M$ is insensitive to the different wave functions used in the
calculation.
In summary, different (but realistic) $\Lambda$ wave functions
give rise to total decay widths which may differ at most by 15\%. 

\subsection{Decay widths of medium-heavy hypernuclei}
\label{heavy}

Using the new W-S wave functions and the Landau parameters $g^{\prime}=0.8$,
$g^{\prime}_{\Lambda}=0.4$ we have extended the 
calculation to heavier hypernuclei. 
We note that, in order to reproduce 
the experimental $s$ and $p$ levels for the hyperon we must
use potentials with nearly 
constant depth, around $28\div 32$ MeV, from medium to heavy hypernuclei
(radii and depths of the used W-S potentials are quoted in table \ref{ws par}).
\begin{table}[t]
\begin{center}
\caption{W-S parameters}
\label{ws par}
\begin{tabular}{c|c c}
\mc {1}{c|}{$ ^{A+1}_{\Lambda}Z$} &
\mc {1}{c}{$R$ (fm)} &
\mc {1}{c}{$V_0$ (MeV)} \\ \hline\hline
$^{12}_{\Lambda}$C    & 2.27 & -32.0 \\
$^{28}_{\Lambda}$Si   & 3.33 & -29.5  \\
$^{40}_{\Lambda}$Ca   & 4.07 & -28.0 \\
$^{56}_{\Lambda}$Fe   & 4.21 & -29.0 \\
$^{89}_{\Lambda}$Y    & 5.07 & -28.5 \\
$^{139}_{\Lambda}$La  & 6.81 & -27.5 \\
$^{208}_{\Lambda}$Pb  & 5.65 & -32.0 \\ 
\end{tabular}
\end{center}
\end{table}

\begin{figure}
\begin{center}
\mbox{\epsfig{file=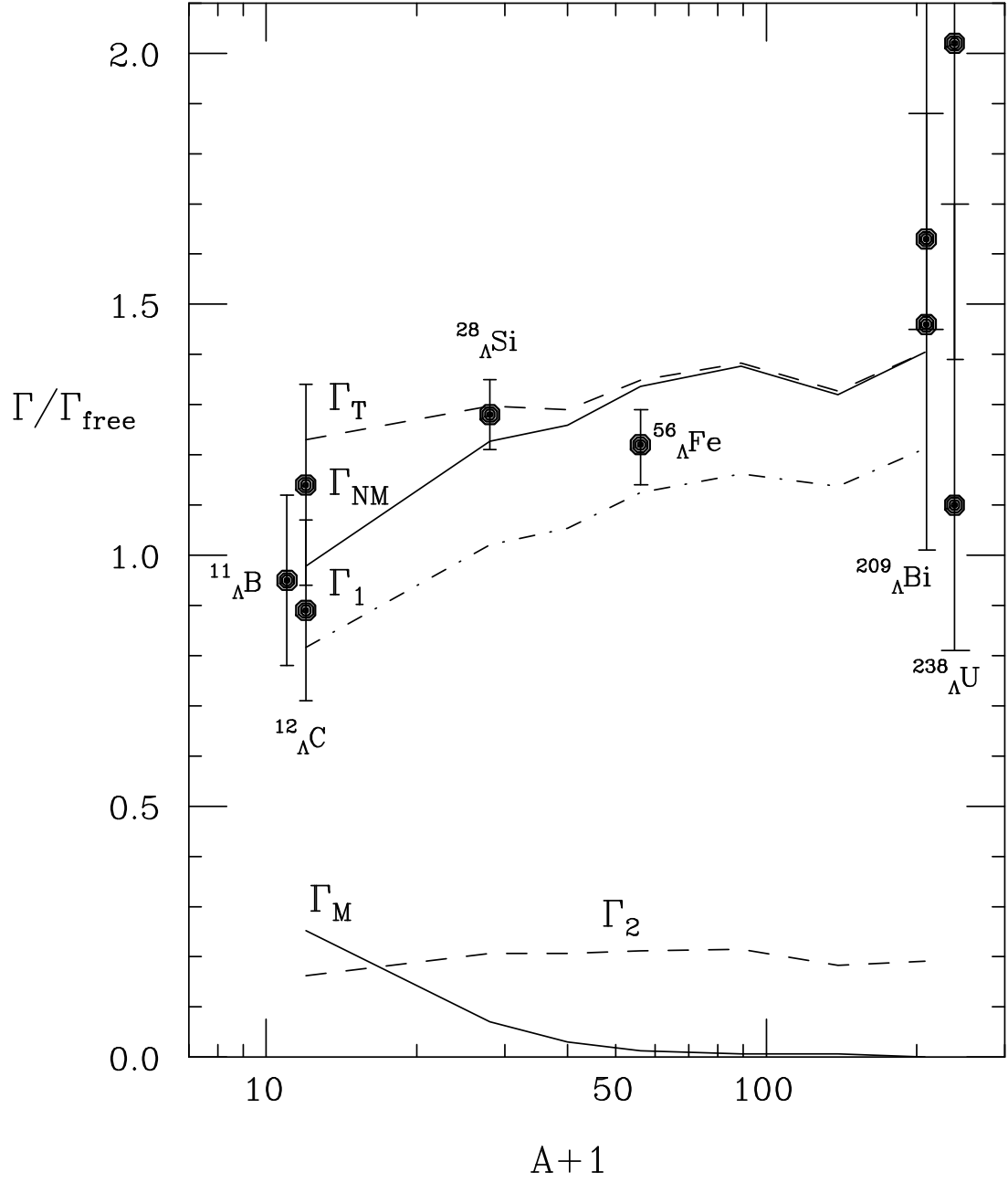,width=.9\textwidth}}
\vskip 2mm
\caption{${\Lambda}$ decay widths in finite nuclei as a function of
the mass number $A$.}
\label{satu}
\end{center}
\end{figure}

Our results are shown in table~\ref{sat}. 
\begin{table}[t]
\begin{center}
\caption{Decay rates}
\label{sat}
\begin{tabular}{c|c c c c}
\mc {1}{c|}{$ ^{A+1}_{\Lambda}Z$} &
\mc {1}{c}{${\Gamma}_M$} &
\mc {1}{c}{${\Gamma}_1$} &
\mc {1}{c}{${\Gamma}_2$} &
\mc {1}{c}{${\Gamma}_{TOT}$} \\ \hline\hline
$ ^{12}_{\Lambda}$C    & 0.25             & 0.82 & 0.16 & 1.23 \\
$ ^{28}_{\Lambda}$Si   & 0.07             & 1.02 & 0.21 & 1.30 \\
$ ^{40}_{\Lambda}$Ca   & 0.03             & 1.05 & 0.21 & 1.29 \\
$ ^{56}_{\Lambda}$Fe   & 0.01             & 1.12 & 0.21 & 1.35 \\
$ ^{89}_{\Lambda}$Y    & $6\cdot 10^{-3}$ & 1.16 & 0.22 & 1.38 \\
$ ^{139}_{\Lambda}$La  & $6\cdot 10^{-3}$ & 1.14 & 0.18 & 1.33 \\
$ ^{208}_{\Lambda}$Pb  & $1\cdot 10^{-4}$ & 1.21 & 0.19 & 1.40 \\
\end{tabular}
\end{center}
\end{table}
We observe that the mesonic rate rapidly vanishes 
by increasing the mass number 
$A$. This is well known and it is related to the decreasing phase space 
allowed for the mesonic channel, and to smaller overlaps between 
the ${\Lambda}$ wave function ${\psi}_{\Lambda}$ and the nuclear surface,
as $A$ increases.
The 2-body induced decay is rather independent of the hypernuclear dimension
and it is about 15\% of the total width. Previous works
\cite{Al91,Ra95} gave more emphasis to this new channel,
without, however, reproducing the experimental results.
The total width is also
nearly constant with $A$, as we already know from the experiment.
In fig.~\ref{satu} we compare the results from table~\ref{sat}
with recent (after 1990) experimental data for non-mesonic decay
\cite{Sz91,No95,Bh98,Ku98,Ar93,Oh98}. 
Nevertheless, we remind that
the data for nuclei from $^{28}_{\Lambda}$Si on refer to the total 
width. However, as can be seen from table~\ref{sat}, 
${\Gamma}_M$($^{28}_{\Lambda}$Si)/${\Gamma}_{NM}$(
$^{28}_{\Lambda}$Si) $\simeq 6\cdot
10^{-2}$ and this ratio rapidly decreases with $A$. The theoretical results
are in good agreement with the data (which, on the other hand, have large error 
bars) over the whole hypernuclear mass range explored. Moreover, we also 
see how the saturation of the ${\Lambda}N\rightarrow NN$ interaction 
in nuclei is well reproduced.

One of the open problems in the study of weak hypernuclear decays is to
understand the large experimental value of the ratio $\Gamma_n/\Gamma_p$ which
most of the present theories fail to reproduce. Only the
quark model of ref.~\cite{Ok98} predicts an enhanced ratio, although
it cannot describe both mesonic and non-mesonic decays from
the same basic quark hamiltonian.
However, we have to remind that the data for $\Gamma_n/\Gamma_p$ have a large
uncertainty and they have been analyzed without taking
into account the 3-body decay mechanism. 
The study of ref.~\cite{Ra95} showed that, even if the three body
reaction is only about 15 \% of the total decay rate, this mechanism 
influences the analysis of the data determining the ratio $\Gamma_n/\Gamma_p$.
The energy spectra of neutrons and protons from the non-mesonic decay
mechanisms were calculated in ref. \cite{Ra97}. The momentum distributions of
the primary nucleons were determined from the Propagator Method 
and a subsequent Monte Carlo simulation was used to account for the final state
interactions. It was shown that the shape of the proton spectrum was sensitive to
the ratio $\Gamma_n/\Gamma_p$. In fact, the protons from the three-nucleon
mechanism appeared mainly at low energies, while those from the
two-body process peaked around 75 MeV. Since the experimental spectra show a
fair
amount of protons in the low energy region they would favour a relatively larger
three-body decay rate or, conversely, a reduced number of protons from the
two-body process. 
Consequently, the experimental spectra are compatible with
values for $\Gamma_n/\Gamma_p$ around $2 \div 3$, in strong contradiction with
the present theories.

\begin{figure}
\begin{center}
\mbox{\epsfig{file=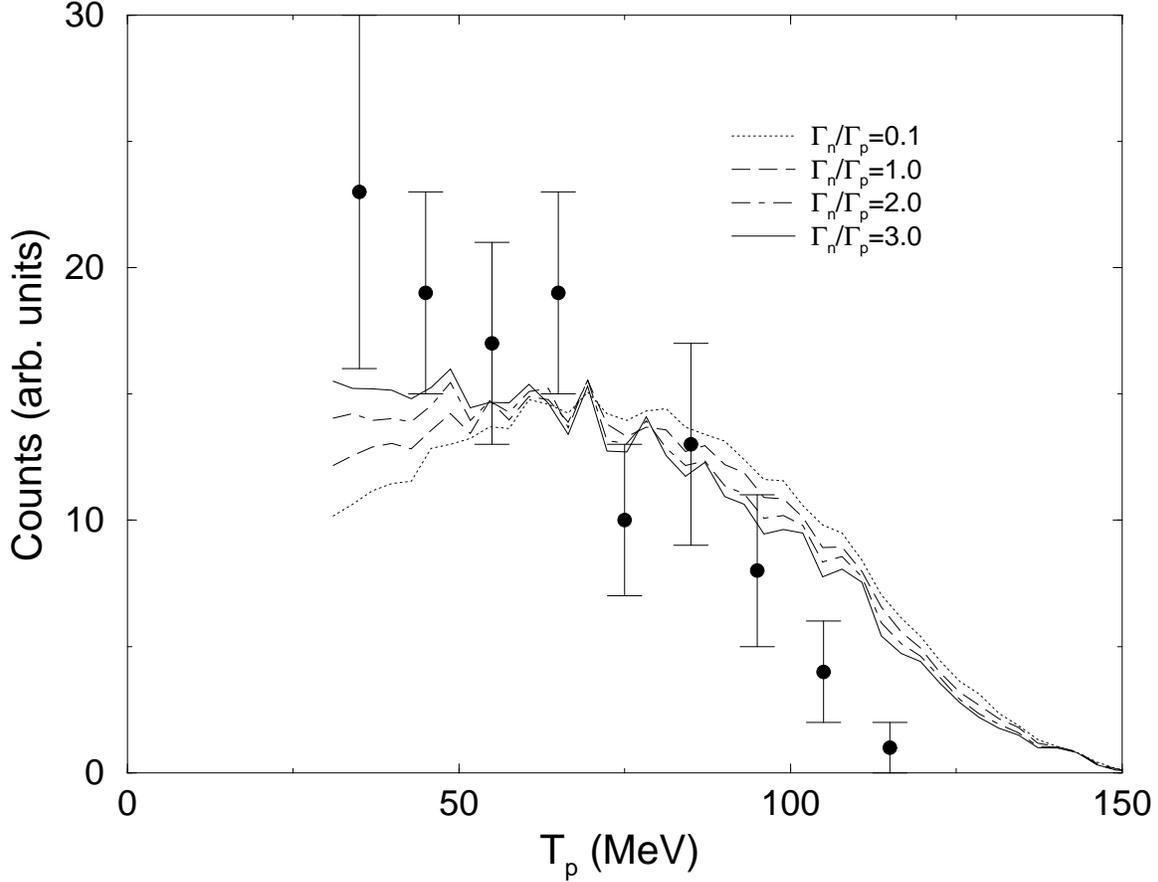}}
\vskip 2mm
\caption{
Proton spectrum from the decay of $^{12}_\Lambda$C for various values
of $\Gamma_n/\Gamma_p$. The experimental data are taken from 
ref.~\protect\cite{Sz91}
}
\label{fig:spec}
\end{center}
\end{figure}

The excellent agreement with the experimental decay rates of medium to heavy
hypernuclei obtained here from the Propagator Method with modified parameters,
makes it worth to explore the predictions for the nucleon spectra.
The question is whether this modified model affects the momentum distribution of
the primary emitted nucleons strongly enough, such that good
agreement with
the experimental proton spectra is obtained without the need for
very large values for $\Gamma_n/\Gamma_p$. We have thus generated
the nucleon spectra from the decay of several hypernuclei
using the Monte Carlo simulation of ref.~\cite{Ra97} but with
our modified $g^{\prime}$, $g^{\prime}_\Lambda$ parameters and our more
realistic nuclear
density and $\Lambda$ wave functions. 
The spectra obtained for various values of $\Gamma_n/\Gamma_p$, used as a 
free parameter in the approach of ref.~\cite{Ra97}, are compared
with the BNL experimental data \cite{Sz91} in fig.~\ref{fig:spec}.
We observe that, although the non-mesonic widths are smaller 
by about 35\% than those of refs.~\cite{Ra95,Ra97}, the resulting nucleon
spectra,
once they are normalized to the
same total width, are practically identical. 
The reason is that the ratio $\Gamma_2/\Gamma_1$ of
two-body induced versus one-body induced decay rates is
essentially the same in
both models (between
0.2 and 0.15 from medium to heavy hypernuclei), and the 
momentum distributions for the primary emitted protons are also very similar.
As a consequence, the conclusions drawn in ref.~\cite{Ra97} still hold {and the
present calculation would
also favour very large values of $\Gamma_n/\Gamma_p$.}

Therefore, the
origin of the discrepancy between theory and experiment for the ratio
$\Gamma_n/\Gamma_p$ still
needs to be resolved. From the theoretical side, there is still room for
improving on the numerical simulation of final state interactions. 
In particular, Coulomb distorsions and the evaporating processes
need to be incorporated. We think that the
evaporating process is an important ingredient which increases the 
nucleon spectra at low energies. Maybe this contribution is so important 
that there is no need for high ${\Gamma}_n/{\Gamma}_p$ values.
On the experimental side,
although new spectra are now available \cite{No95,Bh98}, they have
not been corrected for energy losses inside the target or detector, so a
direct comparison with the theoretical predictions is not yet possible.
Attempts to incorporate these corrections by combining a theoretical model 
for the nucleon rescattering in the nucleus with a simulation of the energy 
losses in the experimental set-up are now being pursued \cite{Ou99}. These
efforts call for newer improved theoretical models that incorporate those  
final state interaction effects missing in ref.~\cite{Ra97}. On the other
hand, a forward step towards
a clean extraction of the ratio
$\Gamma_n/\Gamma_p$ would be obtained if
the nucleons from the different
non-mesonic processes, $\Lambda N \to NN $ and $\Lambda NN \to NNN$ were
disentagled.
Through the measurement 
of the coincident spectra of the outgoing nucleons, it could be possible, in the
near future, to
split the non-mesonic decay width into its two components
${\Gamma}_1$ and ${\Gamma}_2$ \cite{Ze98} and obtain a cleaner measurement 
of the ratio $\Gamma_n/\Gamma_p$.

\section{Conclusions}
\label{concl}
Using the Propagator Method in Local Density Approximation, 
in this paper we made a new evaluation of the ${\Lambda}$ decay 
widths in nuclei. Special attention has been devoted to the study of the
$NN$ and ${\Lambda}N$ short range interactions and realistic nuclear densities
and ${\Lambda}$ wave functions were used. We have adjusted the parameters that 
control the short range correlations to reproduce the experimental decay widths 
of $^{12}_{\Lambda}$C. Then, the calculation has been extended to heavier 
hypernuclei, up to $^{208}_{\Lambda}$Pb. We reproduce for the first time the
experimental non-mesonic decay widths from medium to 
heavy ${\Lambda}$-hypernuclei and the saturation of the
$\Lambda N \to NN$ interaction is observed. 

The energetic spectra of emitted nucleons calculated using 
the Propagator Method with modified parameters (describing the energy 
distributions of primary nucleons) and the Monte Carlo simulation 
(accounting for the final state interactions) does not change appreciably with 
respect to those calculated in ref.~\cite{Ra97}. The reason is that, in spite 
of the fact that the
non-mesonic decay widths ${\Gamma}_1$ and ${\Gamma}_2$ are 
sizably reduced (by about 35\%) with
respect to those of ref.~\cite{Ra97}, the ratio ${\Gamma_1}/{\Gamma_2}$ 
is not altered,
and the momentum distributions of primary nucleons are very similar
to the previous calculation. So, the conclusion
drawn in ref.~\cite{Ra97} still holds: a comparision of the calculated spectra 
with the experimental one favours ${\Gamma}_n/{\Gamma}_p$ ratios around 
$2\div 3$ (or higher), in disagreement with the OPE predictions.
On the other 
hand, we have to recall that for a clean experimental extraction of the 
${\Gamma}_n/{\Gamma}_p$ ratio it is very important to identify 
the nucleons which come out from the different
non-mesonic processes \cite{Ag98}.

\acknowledgments

We would like to thank H. Noumi and H. Outa for discussions and
for giving us detailed information about the experiments. We acknowledge
financial support from the EU contract CHRX-CT 93-0323. This work is also
supported by the MURST (Italy) and the DGICYT contract number PB95-1249 (Spain).

\appendix 
\section*{Spin-isospin $NN$ and ${\Lambda}N\rightarrow NN$ 
interactions}
\label{appen}

In this appendix we show how the repulsive $NN$ and ${\Lambda}N$ Short Range 
Correlations 
(SRC) are implemented in the $NN\rightarrow NN$ and 
${\Lambda}N\rightarrow NN$ interactions.
The process $NN\rightarrow NN$ can be described through an effective potentials
given by:
\beq
\label{gmat}
G(r)=g(r)V(r) .
\eeq 
Here $g(r)$ is a 2-body correlation function, which vanishes as $r\rightarrow 0$
and goes to 1 as $r\rightarrow \infty$, while $V(r)$ is a meson exchange
potential which in our case
contains $\pi$ and $\rho$ exchange: $V=V_{\pi}+V_{\rho}$. 
A practical form for $g(r)$ is \cite{Os82}:
\beq
\label{pract}
g(r)=1-j_0(q_cr) ,
\eeq
With $q_c\simeq 780$ MeV one get a good reproduction of realistic
$NN$ correlation functions obtained from $G$-matrix calculations. 
The inverse of $q_c$ is indicative of the hard core radius of the interaction.
Since there are not experimental indications, 
the same correlation momentum we use for the ${\Lambda}N$ interaction.
On the other hand, we remind that $q_c$
is not necessarily the same in the two cases, given the different nature of
the repulsive forces. Using the correlation function (\ref{pract}) it is easy to
get the effective interaction, eq.~(\ref{gmat}), in momentum space. 
It reads:
\beq
\label{spinisospin}
G_{NN}(q)=V_{\pi}(q)+V_{\rho}(q)+
\frac{f_{\pi}^2}{m_{\pi}^2}\left\{g_L(q)\hat{q}_i \hat{q}_j+
g_T(q)({\delta}_{ij}-\hat{q}_i \hat{q}_j)\right\}{\sigma}_i{\sigma}_j
\bbox{\tau}\cdot \bbox{\tau} ,
\eeq
where the SRC are embodied in the correlation functions $g_L$ and $g_T$.
The spin-isospin $NN\rightarrow NN$ interaction can be separated into a
spin-longitudinal and a spin-transverse parts, as follows:
\beq
\label{gnn}
G_{NN}(q)=\left\{V_L(q)\hat{q}_i \hat{q}_j+V_T(q)({\delta}_{ij}-
\hat{q}_i \hat{q}_j)\right\}{\sigma}_i{\sigma}_j\bbox{\tau}\cdot \bbox{\tau} 
\hspace{0.5in} (\hat{q}_i=q_i/\mid \bbox q\mid) ,
\eeq
where 
\beq
\label{vl}
V_L(q)=\frac{f_{\pi}^2}{m_{\pi}^2}\left\{\bbox q^2F_{\pi}^2(q)G_{\pi}^0(q)+
g_L(q)\right\} ,
\eeq
\beq
\label{vt}
V_T(q)=\frac{f_{\pi}^2}{m_{\pi}^2}\left\{\bbox q^2C_{\rho}F_{\rho}^2(q)
G_{\rho}^0(q)+g_T(q)\right\} .
\eeq
In the above $F_{\rho}$ is the $\rho NN$ form factor (eq.~(\ref{ff})
with cut-off ${\Lambda}_{\rho}=2.5$ GeV), and 
$G_{\rho}^0=1/(q_0^2-\bbox q^2-m_{\rho}^2)$ is the ${\rho}$
free propagator.

The ${\Lambda}N\rightarrow NN$ effective interactions splits into a 
$p$-wave (again longitudinal and transverse) part:
\beq
\label{src}
G_{{\Lambda}N\rightarrow NN}(q)=
\left\{\tilde{P}_L(q)\hat{q}_i \hat{q}_j+\tilde{P}_T(q)({\delta}_{ij}-
\hat{q}_i \hat{q}_j)\right\} {\sigma}_i{\sigma}_j\bbox{\tau}\cdot \bbox{\tau} ,
\eeq
with:
\beq
\label{pl}
\tilde{P}_L(q)=\frac{f_{\pi}}{m_{\pi}}\frac{P}{m_{\pi}}
\left\{\bbox q^2F_{\pi}^2(q)G_{\pi}^0(q)+g_L^{\Lambda}(q)\right\} ,
\eeq
\beq
\label{pt}
\tilde{P}_T(q)=\frac{f_{\pi}}{m_{\pi}}\frac{P}{m_{\pi}}g_T^{\Lambda}(q) ,
\eeq
and an $s$-wave part:
\beq
\label{s}
\tilde{S}(q)=\frac{f_{\pi}}{m_{\pi}}S\left\{F_{\pi}^2(q)G_{\pi}^0(q)-
\tilde{F}_{\pi}^2(q)\tilde{G}_{\pi}^0(q)\right\}\mid \bbox q \mid .
\eeq
Form factors and propagators with a tilde imply that they are calculated
changing $\bbox q^2\rightarrow~\bbox q^2+q_c^2$. $C_{\rho}$ is given by the
expression:
\beq
C_{\rho}={\frac{f_{\rho}^2}{m_{\rho}^2}}
\left[\frac{f_{\pi}^2}{m_{\pi}^2}\right]^{-1} .
\eeq
The expressions for the correlation functions are the following:
\beq
\label{gl}
g_L(q)=-\left\{\bbox q^2+\frac{1}{3}q_c^2\right\}\tilde{F}_{\pi}^2(q)
\tilde{G}_{\pi}^0(q)-\frac{2}{3}q_c^2C_{\rho}\tilde{F}_{\rho}^2(q)
\tilde{G}_{\rho}^0(q) ,
\eeq
\beq
\label{gt}
g_T(q)=-\frac{1}{3}q_c^2\tilde{F}_{\pi}^2(q)\tilde{G}_{\pi}^0(q)
-\left\{\bbox q^2+\frac{2}{3}q_c^2\right\}C_{\rho}\tilde{F}_{\rho}^2(q)
\tilde{G}_{\rho}^0(q) ,
\eeq
\vspace{0.08in}
\beq
\label{gll}
g_L^{\Lambda}(q)=-\left\{\bbox q^2+\frac{1}{3}q_c^2\right\}\tilde{F}_{\pi}^2(q)
\tilde{G}_{\pi}^0(q) ,
\eeq
\beq
\label{gtl}
g_T^{\Lambda}(q)=-\frac{1}{3}q_c^2\tilde{F}_{\pi}^2(q)\tilde{G}_{\pi}^0(q) .
\eeq
Using the set of parameters:
\beq
\label{parameters}
q_c=780\, {\rm MeV}, \hspace{0.1in} {\Lambda}_{\pi}=1.2\,
{\rm GeV}, \hspace{0.1in} 
{\Lambda}_{\rho}=2.5\, {\rm GeV}, \hspace{0.1in} f^2_{\pi}/4\pi= 0.08,
\hspace{0.1in} C_{\rho}=2 , 
\eeq 
at zero energy and momentum we have:
\beq
g_L(0)=g_T(0)=0.615, \hspace{0.4in} 
g_L^{\Lambda}(0)=g_T^{\Lambda}(0)=0.155 .
\eeq
Howhever we wish to keep the zero energy and momentum limit of $g_{L,T}$ 
and $g_{L,T}^{\Lambda}$ as free
parameters, thus we replace the previous functions by:
\beq
g_{L,T}(q)\rightarrow g^{\prime}\frac{g_{L,T}(q)}{g_{L,T}(0)}, \hspace{0.4in}
g_{L,T}^{\Lambda}(q)\rightarrow g^{\prime}_{\Lambda}\frac{g_{L,T}^{\Lambda}(q)}
{g_{L,T}^{\Lambda}(0)} .
\eeq

\vfill\eject

\end{document}